\documentclass[aps,prb,twocolumn,superscriptaddress]{revtex4-1}
\usepackage{graphicx,bm,amssymb,amsmath,xcolor,pifont,soul}
\usepackage[linktocpage=true,colorlinks=true,pdfborder={0 0 0},linkcolor=blue,citecolor=red,filecolor=yellow,urlcolor=blue,bookmarks,pdfauthor={},]{hyperref}
\usepackage{ulem}
\usepackage{multirow}

\def\BiPd{$\beta$-Bi$_2$Pd }
\def\BiPdP{$\beta$-Bi$_2$Pd}

\begin{document}

\title{Microscopic Scattering Approach to In-Gap States: Cr Adatoms on Superconducting $\beta$-Bi$_2$Pd}
\author{Joseph Sink}
\affiliation{Department of Physics and Astronomy, University of Iowa, Iowa City, Iowa 52242, USA}

\author{Hari Paudyal}
\affiliation{Department of Physics and Astronomy, University of Iowa, Iowa City, Iowa 52242, USA}

\author{Divya Jyoti}
\affiliation{Centro de F{\'{i}}sica de Materiales CFM/MPC (CSIC-UPV/EHU),  20018 Donostia-San Sebasti\'an, Spain}
\affiliation{Donostia International Physics Center (DIPC),  20018 Donostia-San Sebasti\'an, Spain}

\author{Andreas J. Heinrich}
\affiliation{Center for Quantum Nanoscience (QNS), Institute for Basic Science (IBS),Seoul 03760, South Korea}
\affiliation{Department of Physics, Ewha Womans University,Seoul 03760, South Korea}

\author{Deung-Jang Choi}
\affiliation{Centro de F{\'{i}}sica de Materiales CFM/MPC (CSIC-UPV/EHU),  20018 Donostia-San Sebasti\'an, Spain}
\affiliation{Donostia International Physics Center (DIPC),  20018 Donostia-San Sebasti\'an, Spain}
\affiliation{Ikerbasque, Basque Foundation for Science, 48013 Bilbao, Spain}

\author{Nicol{\'a}s Lorente}
\affiliation{Centro de F{\'{i}}sica de Materiales CFM/MPC (CSIC-UPV/EHU),  20018 Donostia-San Sebasti\'an, Spain}
\affiliation{Donostia International Physics Center (DIPC),  20018 Donostia-San Sebasti\'an, Spain}

\author{Michael E. Flatt\'e}
\affiliation{Department of Physics and Astronomy, University of Iowa, Iowa City, Iowa 52242, USA}
\affiliation{Department of Applied Physics and Science Education, Eindhoven University of Technology, Eindhoven, The Netherlands}

\date{\today}

\begin{abstract}
We develop a microscopic scattering formalism to describe Yu-Shiba-Rusinov (YSR) states due to a single Cr adatom on the Bi-terminated surface of $\beta$-Bi$_2$Pd, by combining \textit{ab initio} Wannier functions with a real-space Green’s function approach in the Bogoliubov–de Gennes formalism.
Our framework reproduces key scanning tunneling spectroscopy features, including a single particle-hole asymmetric YSR peak and isotropic $dI/dV$ maps around the impurity. 
Decomposing the YSR states reveals contributions from four nearly degenerate $C_{4v}$ representations, with energy broadening masking their individual signatures. Spin-orbit coupling induces partial spin polarization, while the spatial asymmetry between particle and hole components arises from Cr $d$–Bi $p$ hybridization. 
These results highlight the importance of realistic band structures and microscopic modeling for interpreting STM data and provide a foundation for studying impurity chains hosting topological excitations.
\end{abstract}

\maketitle 

\section{Introduction}

The scanning tunneling microscope (STM) provides unparalleled control at the atomic scale, enabling transformative studies of matter atom-by-atom. Early applications demonstrated the manipulation of individual atoms, revealing the intricate interplay of atomic-scale interactions and paving the way for the study of emergent electronic phenomena at the nanoscale~\cite{Eigler1990,Stroscio1991,Crommie1993,Meyer1996,Morgenstern2013}. These interactions can strongly influence the electronic properties of host materials, giving rise to novel quantum states~\cite{Khajetoorians,Drost2017,Lado}.

A particularly striking example is the behavior of magnetic atoms on superconducting substrates. Such atoms induce localized in-gap bound states, known as Yu-Shiba-Rusinov (YSR) states~\cite{Yu_1965,Shiba_1968,Rusinov_1969}, due to the exchange interaction with the Cooper pairs. These states correspond to spin-polarized excitations within the superconducting gap and reflect the local breaking of time-reversal symmetry and the weakening of the superconducting order parameter near the impurity site.

In recent years, YSR states have attracted renewed interest owing to theoretical proposals and experimental evidence suggesting that chains of magnetic atoms on superconductors can host topological edge states~\cite{Beenakker,Yazdani2013,choi2021,Mier_2022,Mier2021,Jyoti2024}. These edge states, often appearing as zero-bias anomalies in STM spectra, have been identified as potential realizations of Majorana bound states (MBSs) --- quasiparticles with non-Abelian exchange statistics, central to the concept of topological quantum computation~\cite{Simon2008,Alicea_2011,Leijnse_2012}. 

STM has proven indispensable in probing these phenomena, enabling both spectroscopic and spatial mapping of in-gap states~\cite{Yazdani,Pawlak_2016,Yazdani2017,Kim_2018,Schneider_2022,Rachel2025}. In particular, STM-based measurements have revealed detailed orbital and spatial structure of YSR states~\cite{ruby2016,choi2017,choi2018,Schneider1,Liebhaber2022}, providing key insights into the nature of the spin–superconductor interaction. Developing accurate and quantitative theoretical tools to describe the emergence and spatial characteristics of YSR states is therefore crucial for understanding and engineering impurity-induced phenomena in superconductors.

\begin{figure}[ht!]
    \centering
        \includegraphics[width=.9\linewidth]{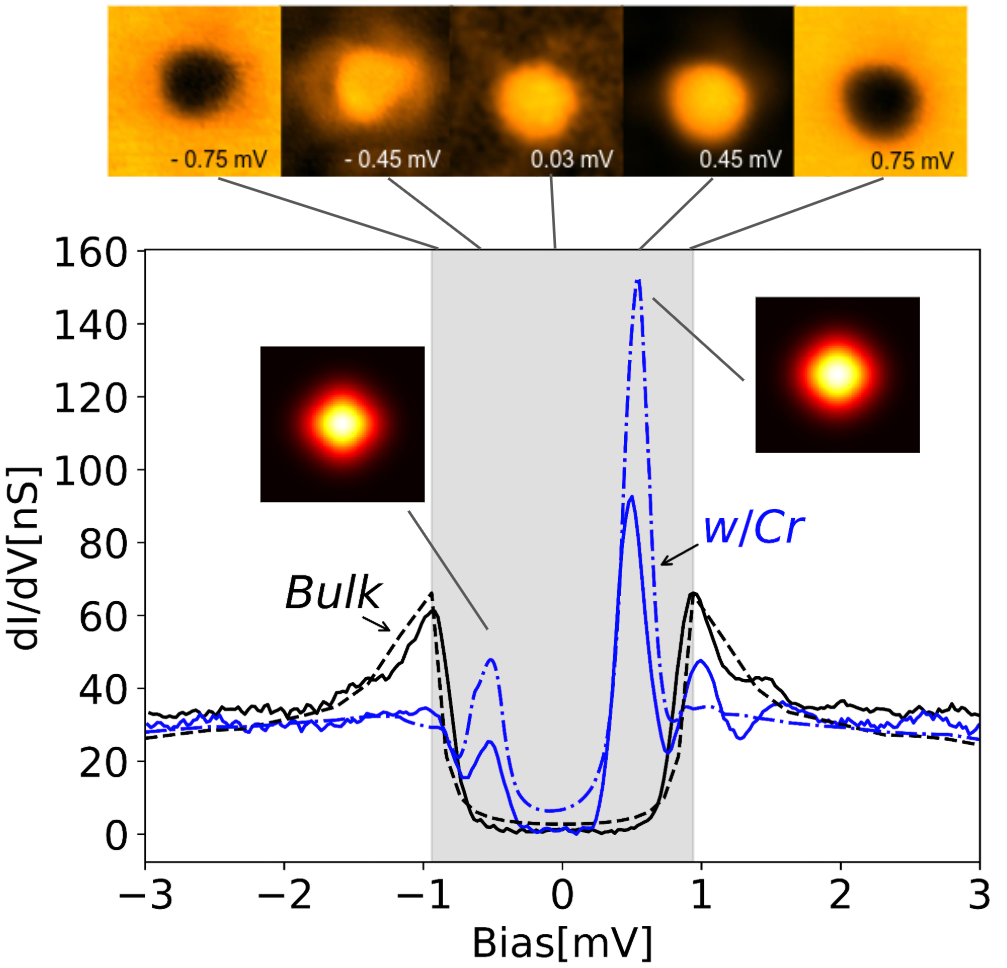}
    \caption{Experimental and theoretical data of in-gap states induced by a single Cr atom on \BiPdP. The upper panel shows the spatial $dI/dV$ map at STM bias voltages corresponding to the superconducting coherence peaks, YSR bound states, and zero bias (which exhibits no distinct orbital character). The lower panel compares the experimental (solid lines) and theoretical (dashed lines) $dI/dV$ spectra for the clean \BiPdP{} surface (black) and over a single Cr atom (blue). Insets show simulated $dI/dV$ maps at the energies of the YSR peaks computed using Eq.~(\ref{eq:STM}). Theoretical spectra are normalized to match the amplitude of the coherence peaks.}
    \label{Fig1}
\end{figure}

Recently, chains of Cr atoms deposited on $\beta$-Bi\(_2\)Pd --- a centrosymmetric type-II superconductor with critical temperature \( T_c = 5.2\,\text{K} \) --- have been shown to induce rich in-gap structures, including possible zero-energy bound states~\cite{choi2021,choi2018}. For isolated Cr atoms, STM measurements reveal a single pronounced YSR peak with substantial particle-hole asymmetry (see Fig.~\ref{Fig1}), in stark contrast to the multiple (up to five) YSR resonances observed for Cr on Pb(111)~\cite{choi2017}. These differences are not well captured by simpler phenomenological models of the superconductor and impurity interaction~\cite{Flatte1997a,FlatteByers.PRB.1997,Salkola_1997,choi2017,choi2018}, and call for a more accurate microscopic treatment.

In this work, we develop a theoretical framework based on real-space Green’s functions derived from the Bogoliubov–de Gennes (BdG) formalism with $s$-wave pairing, used to solve the Gor’kov scattering problem in the presence of a magnetic impurity~\cite{Flatte1997a,FlatteByers.PRB.1997,PhysRevB.61.14810,Flatte1998,Flatte1999b,Csire2015,Saunderson2020,Saunderson2020a,Nyari2021,Park2023}. An accurate description of the normal-state band structure near the Fermi level is essential for this approach\cite{PhysRevB.61.14810,Flatte1998,Flatte1999b,Csire2015,Saunderson2020,Saunderson2020a,Nyari2021,Park2023,Flatte2000,PhysRevB.66.060504,PhysRevB.70.140510}. Here we obtain this description from \textit{ab initio} density-functional tight-binding theory, using maximally localized Wannier functions~\cite{MOSTOFI20142309,Pizzi2019s} to construct the basis. The coupling between the Cr $d$-orbitals and Bi $p$-orbitals is described by a Slater–Koster two-center tight-binding model, with the magnetic impurity treated as a classical spin aligned perpendicular to the surface.

Our theory successfully accounts for the experimentally observed spectral features, including the presence of a single YSR peak and the spatial structure of the in-gap states. These results demonstrate the importance of combining realistic band structure with microscopic impurity modeling to interpret STM data of magnetic atoms on superconductors.

The remainder of this paper is organized as follows. In Section~\ref{sec:exp}, we present the experimental STM and $dI/dV$ spectroscopy data for single Cr atoms adsorbed on the surface of $\beta$-Bi\(_2\)Pd, with spatially and energetically resolved measurements across the superconducting gap. Section~\ref{sec:DFT} describes the first-principles characterization of the system using density functional theory within the PBE approximation, focusing on the band structure and orbital-projected density of states of the clean substrate. These results are used to construct a maximally localized Wannier tight-binding model with orbitals centered at atomic positions, providing an interpretable and accurate representation of the low-energy electronic structure. In Section~\ref{sec:theory}, we introduce our theoretical framework to describe impurity-induced in-gap states, based on a real-space implementation of the Gor’kov equations using the Wannier basis. Section~\ref{sec:results} presents the calculated differential conductance spectra and spatial maps of the in-gap states, and compares them to experimental data. We conclude in Section~\ref{sec:conclusions} with a summary of our findings and their implications for understanding impurity-induced states in superconductors.

\section{Experimental}
\label{sec:exp}

The $\beta$-Bi\(_2\)Pd crystal used in our experiments was synthesized following the procedure described in Ref.~[\onlinecite{Imai2012}]. The specific sample selected for this study exhibited a superconducting transition temperature \(T_c\) of 5.2 K, confirming its suitability for low-temperature scanning tunneling microscopy (STM) investigations. To prepare a clean and well-defined surface for atomic manipulation, the $\beta$-Bi\(_2\)Pd crystal was cleaved in situ under ultra-high vacuum (UHV) conditions, exposing a Bi-terminated surface, as detailed in Ref.~[\onlinecite{choi2018}]. 

To introduce magnetic impurities into the system, individual Cr atoms were deposited onto the freshly cleaved $\beta$-Bi\(_2\)Pd surface while maintaining the substrate at a cryogenic temperature of approximately \(T \approx 20\) K. This controlled deposition process ensured the presence of well-isolated single atoms on the surface, minimizing clustering effects 
\cite{Kawaguchi_2019, Lee_2002, Lin_2012, Marakov_2008} that could complicate spectroscopic measurements.

For the STM spectroscopy measurements, we employed a metallic PtIr tip, which allowed for a direct measurement of the local density of states (LDOS) of the $\beta$-Bi\(_2\)Pd substrate through differential conductance (\(dI/dV\)) spectra \cite{choi2021}. The use of a dilution refrigerator \cite{Kim2021} STM at \(T = 30\) mK provided a substantial advantage by ensuring that thermal smearing remained negligible, yielding an energy resolution on the order of a few tens of \(\mu\)eV. This gives us a direct measurement of the LDOS without deconvoluting the superconducting tip used in other studies at higher temperatures~\cite{ruby2016,choi2017,choi2018}.

The region surrounding the Cr impurity was mapped at voltages corresponding to the quasiparticle (QP) peaks, YSR states, and eventual zero-energy states, see Fig. \ref{Fig1}. These $dI/dV$ maps reveal a circular, featureless pattern with no distinct orbital character.
The maps exhibit a clear polarity dependence, indicating electron-hole asymmetry in the wave function of the corresponding YSR state. The slight different shapes of the electron and hole components of the YSR state account for the variation in peak amplitude observed in the point spectra, as seen Fig.~\ref{Fig1}. Furthermore, there are clear in-gap single peaks appearing at positive and negative bias. Given the complex structure of Cr atoms, with an open $d$-shell, the interpretation of previous works~\cite{ruby2016,choi2017} leads to expecting more than one peak.

\section{Theoretical Methods}

\begin{figure*}[ht!]
\begin{center}
   \includegraphics[width=0.99\linewidth] {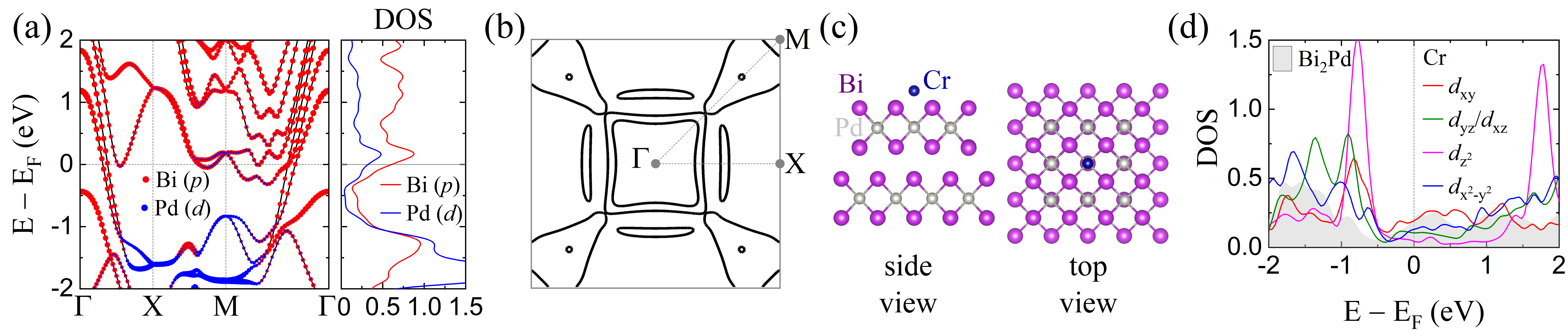}
    \caption{Atomic projected electronic band structure and DOS (/eV/atom) (a) and two-dimensional Fermi surface (b) of two-layers Bi$_2$Pd including SOC. The size of the red and blue circles correspond to the magnitude of the Bi ($p$) and Pd ($d$) contributions, respectively. Crystal structure displaying the side and top views of two-layers Bi$_2$Pd including a Cr atom at the $C_{4v}$ symmetric hollow site (c), and orbital-projected DOS (/eV/atom), highlighting the contributions from Cr $d$-orbitals. The gray-shaded area represents the contribution from Bi$_2$Pd layers.}
    \label{fig-dft}
    \end{center}
\end{figure*}

We model the problem of the single Cr adatom on the surface of \BiPd by solving the real-space Gor'kov equations for a $C_{4v}$ symmetric localized magnetic impurity\cite{Flatte1998,Flatte1999b}. The normal and anomalous homogeneous Green's functions for the pristine \BiPd system are calculated within the Bogoliubov-de Gennes (BdG) formalism, assuming s-wave pairing and realistic normal metallic states.

\subsection{DFT TB: \BiPd}
\label{sec:DFT}

The normal metallic states for \BiPd were computed with spin-orbit coupling (SOC) using \texttt{Quantum ESPRESSO} \cite{giannozzi2017o}. Relativistic norm-conserving pseudopotentials~\cite{PseudoDojo2018} were used with the Perdew-Burke-Ernzerhof exchange-correlation functional within the generalized gradient approximations~\cite{perdew1996o}. Structural relaxation and self-consistent calculations were converged using a kinetic-energy (charge density) cutoff of 80~Ry (320~Ry), a $\Gamma$-centered $16\times16\times1$ Monkhorst-Pack \textbf{k}-mesh~\cite{Monkhors1976s}, and an MP-smearing~\cite{Methfessel1989s} value of 0.01~Ry. 

In the bulk phase, \BiPd crystallizes in a centrosymmetric tetragonal crystal structure with space group $I4/mmm$ (No.~139)~\cite{herrera2015magnetic}. We model the bilayer \BiPd surface by taking two tri-layer sheets (6 atoms deep) from the bulk with 25~\AA~is of vacuum along the $z$-direction in order to prevent interactions between periodic images. The optimized in-plane lattice parameter of \BiPd bilayer is found 3.333~\AA, in good agreement with the experimentally  measured in-plane lattice parameter of the bulk \BiPd~\cite{herrera2015magnetic}. 

Figure~\ref{fig-dft}(a-b) shows electronic band structure, atom projected DOS, and Fermi surface of two-layers \BiPd calculated including SOC. It is observed that four bands, primarily characterized by Bi ($p$)-orbitals, cross the Fermi level. As a result, the DOS at the Fermi level has a 65\% contribution from the Bi ($p$)-orbitals, with the remaining from the Pd ($d$)-orbitals. 

Among the four bands, two hole-like bands centered around $\Gamma$ form square-shaped Fermi surfaces, with the outer corner that merge with an adjacent Fermi pocket along the $\Gamma$-M direction. Additionally, two small electron pockets along $\Gamma$-X and $\Gamma$-M are observed, originating from shallow valleys located slightly above the Fermi level. The calculated Fermi surface agrees well with the experimental Fermi surface obtained by angular-resolved photoemission spectroscopy~\cite{sakano2015}.

Wannier90\cite{Pizzi2019s} was used to transform the delocalized plane-wave solutions into a local orbital basis. This basis is highly advantageous for describing local effects (e.g., perturbations to surface 
 Bi atoms due to the presence of a single magnetic Cr adatom).
The centers of the Wannier orbitals were constrained to coincide with the atomic centers using the {\it Selectively Localized Wannier Function+Centers} (SLWF+C) feature\cite{SLWF}, conveniently modified.\footnote{ 
This function operates by imposing a distance dependent penalty function that in general may result in SLWF+C solutions that are more delocalized than the default maximally localized Wannier functions (MLWF). We found in this instance that the center constraint had negligible impact on the Wannier localization. We used a modified version of Wannier90 in order to allow us to simultaneously constrain all Wannier atomic centers.}

The normal state electronic {\it ab initio} Wannier90 Tight-Binding Hamiltonian, $\hat{H}_0$, is of the form
\begin{align}
\hat{H}_0(\mathbf{k})=\sum_{\beta\beta'}e^{i\mathbf{k}\cdot(\mathbf{r}_\beta-\mathbf{r}_{\beta'})}h_{\beta\beta'}c^\dagger_{\beta}c_{\beta'}
\end{align}
where $\beta=\{\mathbf{R}_{ijk},n,\sigma\}$ is a compound index containing the unit cell $\mathbf{R}_{ijk}=i\mathbf{a}_1+j\mathbf{a}_2+k\mathbf{a}_3$, the Wannier-orbital index $n$ (with center at $\mathbf{r}_n=\langle w_n|\hat{\mathbf{r}}|w_n\rangle$), and the spin index $\sigma$.\footnote{See Supplemental for raw files.}

Our Wannier basis set includes 4 sets of Bi $p$-orbitals, 2 sets of Pd $d$-orbitals, for a total of 44 bands w/SOC. The strength of the SOC ($\zeta$) extracted ($H_{SOC}=\zeta\hat{l}\cdot\hat{s}$) from the Wannierized {\it ab initio} calculation is 1.123~eV and 815~meV for the surface and sub-surface Bi p-orbitals, respectively, and 179~meV for the Pd $d$-orbitals. 

Finally, the adsorption of Cr on the \BiPd was not described in terms of Wannier functions, but rather in atomic orbitals, as explained in the following section.

\subsubsection{DFT: Cr adatom on \BiPd Surface}
\label{sec:theory}

To confirm that Cr favors the $C_{4v}$ symmetric Bi hollow site on \BiPd, we relaxed the geometry for a single Cr atom on the surface of a 3$\times$3$\times$1 supercell. It is found that the most stable configuration is the $C_{4v}$ symmetric bismuth hollow site 0.95~\AA~ above the top of the Bi layer with Bi-Cr bond length of 2.64~\AA. 
The nearest Bi atoms experience an in-plane/out-of-plane distortions of approximately 0.1/0.2~\AA.

The projected DOS for Cr atom on \BiPd is shown in Fig.~\ref{fig-dft}(d).
There is a prominent $d_{z^2}$ peak around -1~eV, indicating a localized state that potentially leads to influence the quasiparticle dynamics. 
The finite $d$-orbital states at the Fermi level suggest possible itinerant magnetism induced by the Cr atom, which may modify the superconductingting pairing mechanism in the limit of large Cr coverage.

A Cr spin moment of 3.7~$\mu_B$ was computed from the self-consistent DFT calculations including SOC indicating a negligible induced orbital moment. This demonstrates that Cr favors a high-spin state that tend to be robust against crystal field effects. The large spin leads to a larger $d$-shell occupation of the majority spin of Cr.

These results were obtained for $U=0$. Previous calculations using GGA+U show an important decrease of the Cr density of states at the Fermi energy~\cite{choi2018} as expected from the effect of the Hubbard $U$ on the Cr $d$-orbitals. While finite $U$ is more realistic to describe the electronic structure of transition-metal impurities (see for example, Ref. \cite{Abufager}), the present calculation explores different parameters showing robust behavior with respect to in-gap states.

\subsection{Scattering theory}
\label{sec:theory}

To describe the spectral and quasiparticle properties of a superconductor in the presence of perturbations such as impurities or interfaces, we start from the Bogoliubov–de Gennes (BdG) formalism. This framework allows a mean-field treatment of the superconducting state, encoding the particle-hole coherence induced by the superconducting condensate.

We define the BdG Hamiltonian in the Nambu spinor basis as follows:
\begin{align}
    \hat{H}_{BdG}(\mathbf{k}) =
    \begin{pmatrix}
        \hat{H}_0(\mathbf{k}) & \hat{\Delta} \\
        \hat{\Delta}^\dagger & -\hat{H}^*_0(-\mathbf{k})
    \end{pmatrix},
    \label{eq:BdGHam}
\end{align}
where the spinor basis is given by $\Psi^\dagger_k = (\psi^\dagger_{k\uparrow}, \psi^\dagger_{k\downarrow})$ for the electron sector and $\overline{\Psi}_{-k} = (\psi_{-k\downarrow}, -\psi_{-k\uparrow})$ for the hole sector. 

Here $\hat{H}_0(\mathbf{k})$ denotes the normal-state Hamiltonian, which includes the kinetic energy, spin-orbit coupling (SOC), and other single-particle terms. The superconducting pairing potential $\hat{\Delta}$ describes the condensate, and is assumed to be spin-singlet and local in momentum space. To enforce the antisymmetry required by Fermi statistics, the pairing matrix must satisfy:
\begin{align}
    \hat{\Delta} = \mathbf{1} \otimes (i \hat{\tau}_2)\, \overline{\Delta},
\end{align}
where $\hat{\tau}_2$ is the second Pauli matrix acting in spin space, and $\overline{\Delta}$ is a scalar amplitude (or more generally a scalar function or matrix in orbital space) representing the gap function. The structure ensures that $\hat{\Delta} = -\hat{\Delta}^T$, in accordance with the singlet pairing symmetry.

To compute the physical observables in real space, particularly the local density of states (LDOS), we first construct the real-space Green’s function of the homogeneous superconducting system. This is obtained by Fourier transforming the resolvent of the BdG Hamiltonian\cite{Flatte1997a,Flatte1999b}:
\begin{align}
    \hat{g}_{BdG}(\delta\mathbf{R}; z) = \int_{\Omega_{BZ}} \left[ z\hat{1} - \hat{H}_{BdG}(\mathbf{k}) \right]^{-1} e^{-i\mathbf{k}\cdot\delta\mathbf{R}}\, d\mathbf{k},
\end{align}
where $z = \omega + i\delta$ is a complex energy variable, with $\omega$ being the real frequency and $\delta$ a positive infinitesimal that accounts for broadening. The integral is performed over the Brillouin zone $\Omega_{BZ}$ using a two-dimensional adaptive h-cubature integration scheme to ensure numerical accuracy.

In the Nambu basis, the \textit{homogeneous} Green's function $\hat{g}_{BdG}(\omega)$ can be written\cite{Flatte1997a,Flatte1999b} in block matrix form as:
\begin{align}
    \hat{g}_{BdG}(\omega) =
    \begin{pmatrix}
        \hat{g}_\sigma(\omega) & \hat{f}_\sigma(\omega) \\
        \hat{f}^*_\sigma(-\omega) & -\hat{g}^*_{-\sigma}(-\omega)
    \end{pmatrix},
    \label{eq:homo}
\end{align}
where $\hat{g}_\sigma(\omega)$ is the normal Green’s function for electrons with spin $\sigma$, and $\hat{f}_\sigma(\omega)$ is the anomalous (Gor’kov) Green’s function that captures the electron-hole pairing correlations. These off-diagonal terms are essential for the emergence of superconductivity and encode the pairing amplitude in the Green’s function formalism.

The quantity of primary interest for comparison with scanning tunneling spectroscopy (STS) experiments is the local density of states (LDOS), which can be extracted from the diagonal (normal) part of the Green’s function. For a given spatial position $\mathbf{R}$ and spin-orbital channel $\alpha$, the LDOS is given by:
\begin{align}
    \eta_\alpha(\mathbf{R}, \pm \omega) = -\frac{1}{\pi} \operatorname{Im} \left[ \operatorname{Tr}_\alpha\, \hat{G}(\mathbf{R}, \mathbf{R}; \omega) \right],
    \label{ldos}
\end{align}
where $\operatorname{Tr}_\alpha$ denotes the partial trace over the orbital and spin degrees of freedom specified by $\alpha$, and the sign $\pm \omega$ distinguishes between the electron ($+\omega$) and hole ($-\omega$) sectors in the Nambu representation. 

Equation~\eqref{ldos} forms the basis for generating simulated STS spectra and cross-sectional $dI/dV$ maps. Specifically, by evaluating $\eta_\alpha(\mathbf{R}, \omega)$ at fixed positions while scanning over $\omega$, we simulate differential conductance spectra. Conversely, fixing $\omega$ and scanning over position yields spatial LDOS maps that reveal the spatial structure of in-gap states and resonances, such as those associated with Yu-Shiba-Rusinov states or impurity-induced bound states.

This framework provides the foundation for implementing a scattering formalism in which local perturbations (impurities, adsorbates, structural defects) can be incorporated via Dyson’s equation (here the Gor'kov equation) to yield the full Green’s function in the presence of scattering. In addition loss of pairing coherence can be included directly by modifying the anomalous Green's function\cite{PhysRevB.66.060504,PhysRevB.70.140510}.  The following sections build upon this foundation to compute the impurity-modified Green’s functions and derive observable signatures of in-gap states.

\begin{figure}[h!]
\begin{center}
   \includegraphics[width=.85\linewidth]{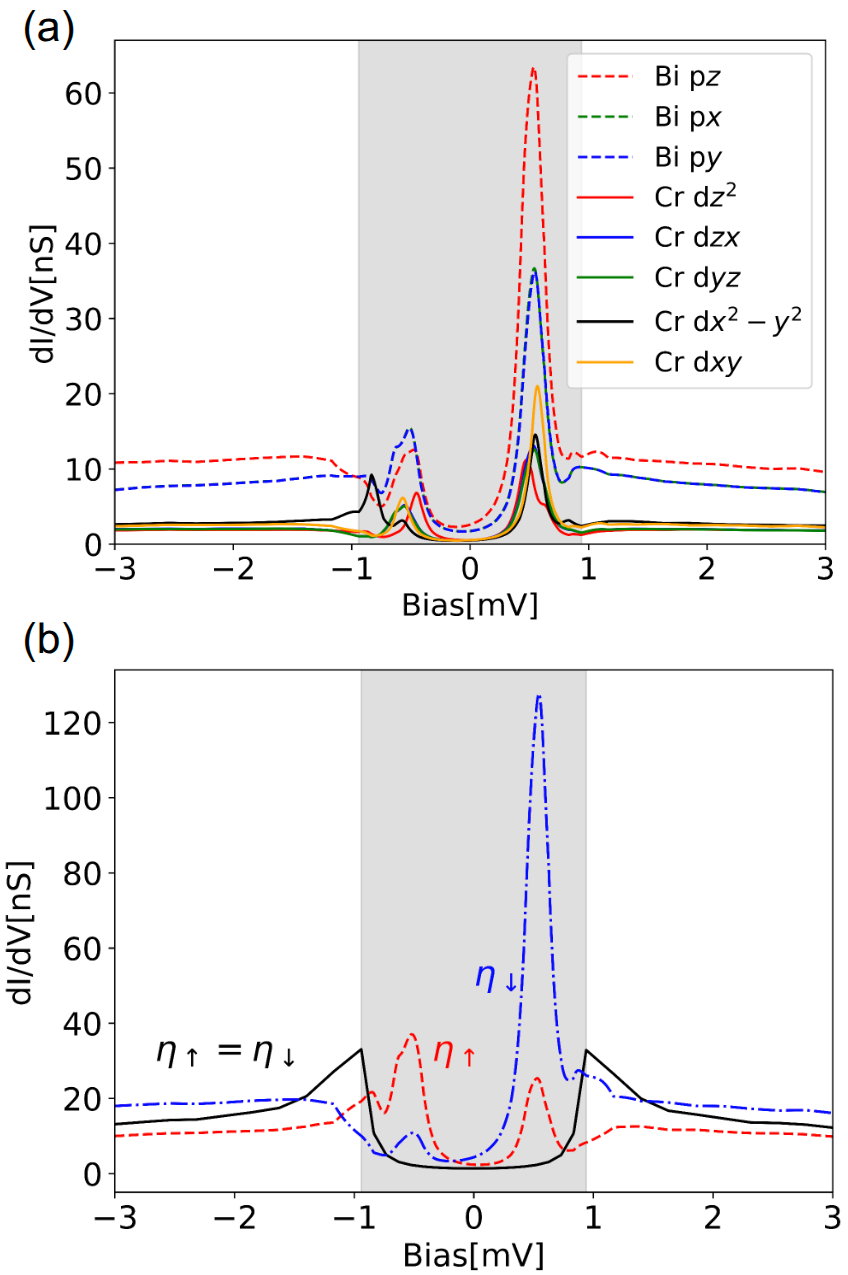} 
    \caption{Spin and orbitally projected local density of states (LDOS). By taking the orbital trace in a), it is apparent that the single peaked feature consists of four separate nearly overlapping YSR states, $A_1\{d_{3z^2-r^2}\}\oplus B_1\{d_{x^2+y^2}\}\oplus B_2\{d_{xy}\}\oplus E\{d_{xz},d_{yz}\}$, identified as the irreducible representations of the $C_{4v}$ point group. The spin partial trace b) was taken by summing over the Cr and Bi nearest neighbor orbitals. This shows the YSR state is partially spin-polarized due to the strong SOC within the Bi layer. Theoretical spectra presented here have been normalized using the same procedure as in Fig.1}
       \label{Fig2}
    \end{center}
\end{figure}

For local perturbations\cite{Flatte1999b}, the potential can be separated into regions where the potential is zero (far-field,f) and non-zero (near-field,n), 
\begin{align}
    \hat{V}'=
    \hat{\tau_z}\otimes\begin{pmatrix}
        \hat{V}_{nn}&0\\
        0&0
    \end{pmatrix},
\end{align}
where $V_{nf}=V_{fn}=V_{ff}=0$ by definition.

The form of the potential allows us to solve the Gor'kov equations exactly \cite{Flatte1997a,FlatteByers.PRB.1997,Flatte1999b}, and yields the inhomogeneous relations,
\begin{align}
    \hat{G}=
    \begin{pmatrix}
        \hat{M}_{nn}\hat{g}_{nn}&\hat{g}_{fn}\hat{M}_{nn}\\
        \hat{M}_{nn}\hat{g}_{nf}&~~~~\hat{g}_{ff}+\hat{g}_{fn}\hat{V}_{nn}\hat{M}_{nn}\hat{g}_{nf}
    \end{pmatrix}\label{eq:Ginhomo}
\end{align}
where $\hat{M}_{nn}=(\hat{1}-\hat{g}_{nn}\hat{V}_{nn})^{-1}$.

The homogeneous solutions, Eq. (\ref{eq:homo}), are computed with open-boundary conditions so that our defect solutions are free from periodicity and finite volume effects. Additionally, from Eq. (\ref{ldos}) and Eq. (\ref{eq:Ginhomo}),  we can see that the inhomogeneous defect solutions, $\hat{G}(\mathbf{R},\mathbf{R}';\omega)$, depend only on the near-field term $\hat{M}_{nn}$ and the homogeneous solution, $\hat{g}_{BdG}$, that connects the two points $\mathbf{R}$ and $\mathbf{R}'$. This allows for the size of the field-of-view of the calculation to be varied without introducing truncation (i.e., boundary or edge) effects. One notable consequence of this feature is that the spectrum of the defect (i.e., simulated STS) can be computed by inverting a small matrix containing only the near-field terms. Simulated cross-sectional dI/dV scans can then be computed at select energies of interest. This offers a great computational advantage over diagonalization-based methods. In particular, the present scattering approach permits us to address the realistic energy scale of the superconducting gap compared to the normal-metal bandwidth, which need to be comparable in diagonalization-based methods.

We model the effect of the Cr impurity as a $C_{4v}$ symmetric interaction between the 4 nearest-neighbor Bi atoms p-states and a classical magnetic moment ($\hat{\mathbf{S}}_{Cr}\rightarrow\langle\hat{\mathbf{S}}_{Cr}\rangle\mathbf{n}$) originating at the Cr,

\begin{align}
    \hat{V}_{nn}=\hat{V}_{Cr}+\hat{V}_{Bi}+\hat{V}_{Cr-Bi},
\end{align}
where the terms describe the atomic Cr d-shell electrons, the the local perturbation to the nearest neighbor chemical potential due to charge transfer with the Cr atom, and the Cr-Bi electronic and exchange coupling, respectively.

The model for the Cr atom used is,
\begin{align}
    \hat{V}_{Cr}=E_d\hat{1}+\lambda \hat{L}\otimes\hat{S}+J_{1}\langle \hat{\mathbf{S}}_{Cr}\rangle\mathbf{n}\cdot\mathbf{\hat{S}}_{Cr},
\end{align}
where the intra-atomic exchange is modelled by a local classical Zeeman-like term 
$J_{1}\langle \hat{\mathbf{S}}_{Cr}\rangle=4$ eV, $\lambda$ is the d-shell spin-orbit taken to be $\lambda=20$ meV, and the center of mass of the d-shell minifold is $E_d-E_{Fermi}=+0.5$ eV. 

The nearest neighbor Bi p-orbitals are offset by $\delta Ep=250$ meV ($\hat{V}_{Bi}=\delta E_p\hat{1}$).
The Cr-Bi electronic and exchange coupling model is,
\begin{align}
    \hat{V}_{Bi-Cr}=
    (\sum_{i}t_{i,Cr}~\hat{p}^\dagger_{i}\hat{d}_{Cr}+C.C.)+J_{2}\langle \hat{\mathbf{S}}_{Cr}\rangle\mathbf{n}\cdot\hat{\mathbf{S}}_{Bi}^i
\end{align}
where $J_{2}\langle\hat{\mathbf{S}}_{Cr}\rangle=0.9$ eV  is the inter-atomic exchange coupling between Cr d-shell and Bi p-shell electrons. We use the Slater-Koster two-center model to describe the non-magnetic electronic hybridization, $t_{i,Cr}$ of the Cr and Bi atoms ($pd\sigma=0.75$ eV and $pd\pi=0.5$ eV; See Supplemental Material for the explicit form of $t_{i,Cr}$). The orientation of the Cr classical moment, $\mathbf{n}$, is taking to be out of plane.

\subsection{Theoretical results}
\label{sec:results} 

Our homogeneous calculation, solution of Eq. (\ref{eq:homo}) using the atomic Wannier basis set (black dashed line Fig. \ref{Fig1}), exhibits quasiparticle peaks (QP) and a gap match well with the observed bulk gap and DOS observed in experimental $dI/dV$ measured far from the Cr impurity (solid black line Fig. \ref{Fig1}). We use a Dynes broadening ($\delta$ in Eq. \ref{eq:homo}) in the calculation of $\delta=0.07$ meV which sets the maximum spectral resolution. Despite the generally excellent agreement between the theory and experiment, our theory does not capture the Fano-like resonance for energies above the QP energies, black line in Fig. \ref{Fig1}. These features are likely due to the substantial electron-phonon coupling of the \BiPd surface \cite{Margine}. However, the multiband and single-gap features of the \BiPd surface as detected experimentally \cite{herrera2015magnetic} are reproduced within our treatment.

The $dI/dV$ obtained on the Cr atom is shown in Fig.\ref{Fig1} (blue solid line). It exhibits two prominent YSR states and isotropic $dI/dV$ maps near the Cr atom with no clear orbital character. This is in contrast to previous works with Cr on Pb \cite{choi2017} in which all 5 Cr orbitals were clearly visible as distinct YSR states. Our theory permits us to explain this apparent contradiction.

The coupling with the surface proximitizes the Cr $d$-electrons and splits them into 4 YSR states according to the $C_{4v}$ symmetric crystal field compatibility relations. We have plotted the contribution to the YSR states of the different orbitals of the $d$-shell of Cr as continuous lines in Fig. \ref{Fig2} (a). These states can be identified by their orbital character as $A_1\{d_{3z^2-r^2}\}\oplus B_1\{d_{x^2-y^2}\}\oplus B_2\{d_{xy}\}\oplus E\{d_{xz},d_{yz}\}$ as corresponds to the $C_{4v}$ symmetry of the adsorbed Cr atom on the hollow site of the Bi-terminated surface. Figure \ref{Fig2} (a) shows the contribution of each of these symmetry characters to the LDOS. The orbital decomposition shows that the splitting of these states is less than the Dynes broadening. This effective degeneracy of the YSR states leads to the detection of a single peak at the Cr site, in agreement with the experimental Cr $dI/dV$.

Furthermore, Fig. \ref{Fig2} (a) also shows the contribution of Bi orbitals to the YSR states. In dashed lines, the contribution of the $p_z$, $p_x$ and $p_y$ is depicted. The $C_{4v}$ symmetry implies that the contribution of the $p_x$ and $p_y$ orbitals (in the surface plane) are degenerate for each of the four Bi atoms surrounding the Cr impurity. The $p_z$ orbital shows the largest overlap with the Cr atom, and the largest contribution to the unoccupied YSR states. However, for the occupied YSR state the $p_x$ and $p_y$ contribution is larger. This behavior can be correlated to the larger contribution of $d_{z^2}$, $d_{xz}$ and $d_{yx}$ Cr orbitals to the unoccupied YSR states, while the $d_{xy}$ and $d_{x^2-y^2}$ prevail for the occupied in-gap states.

Figure \ref{Fig2} (b) analyzes the results of our calculated LDOS  by computing the spin contribution (blue for spin down or majority spin and red for spin up or minority spin as fixed by the impurity's spin), Eq. (\ref{ldos}). The same is done on a distant Bi atom that shows no spin polarization (black curve). The large spin polarization of YSR states\cite{Wiebe} is clearly recovered. However, the large Rashba-SOC at the \BiPd surface leads to a substantial contribution of both spins for the positive- and negative-energy YSR peaks as opposed to the 100\% spin polarization of YSR  states in Fe-based superconductors \cite{Wiebe}.

\begin{figure}[]
    \centering
       \includegraphics[width=1\linewidth]{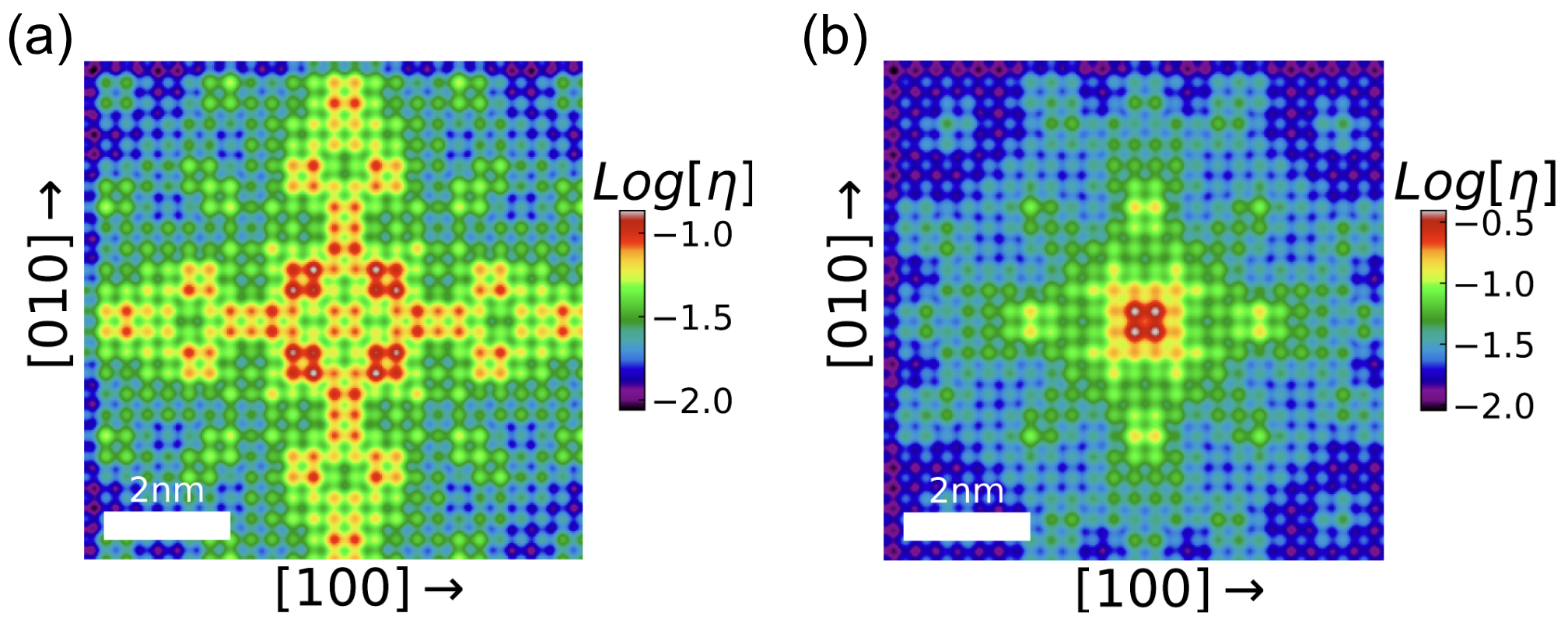}\\
    \caption{Simulated cross-sectional dI/dV for the a) particle contribution to the  YSR state at $\omega=-0.47$ meV and b) hole contribution to the  YSR state at $\omega=0.47$ meV taken in the plane of the surface Bi atoms. The Cr sits above the surface plane. }
    \label{fig4}
\end{figure}

\begin{figure*}[]
    \centering
     \includegraphics[width=.8\linewidth]{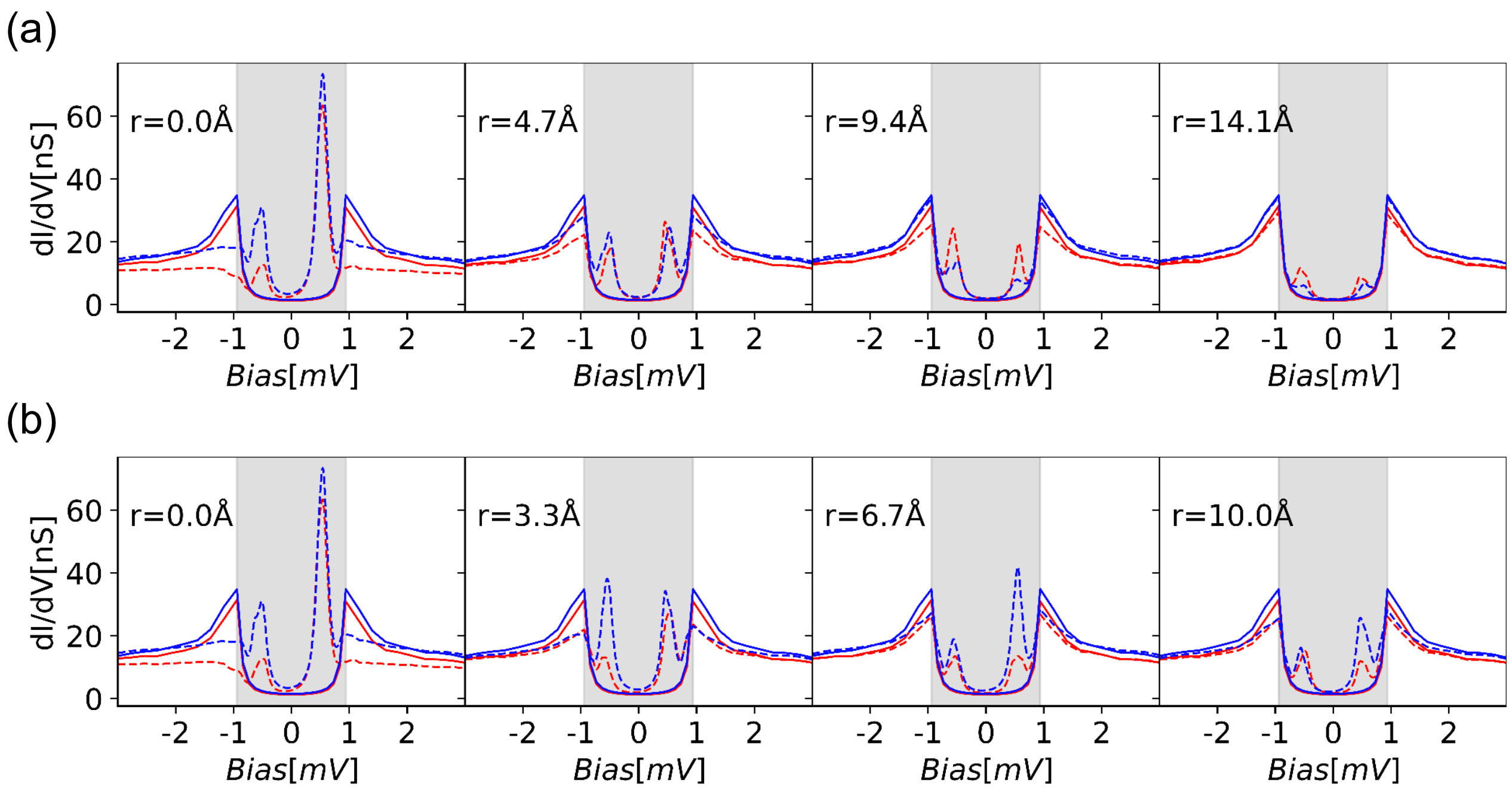}\\
    \caption{Simulated dI/dV taken in the surface plane as a function of distance along the a)$[110]$ and b)$[100]$ directions measured from the nearest-neighbor Bi atom to the Cr impurity. The solid lines and dashed lines correspond the bulk and impurity calculations, respectively. Red and blue represent the partial trace over the $p_z$ and $p_x+p_y$ Bi orbitals, respectively.}
    \label{fig5}
\end{figure*}
The isotropic $dI/dV$ map of the region near the Cr atom, Fig. \ref{Fig1}, is explained similarly. The STM response to the surface is proportional to the join density of states the tip state, $\phi_{tip}(\bf{r})$ and surface state, $\psi$,
\begin{align}
\eta(\mathbf{r})&=\int |\phi^*_{tip}(\mathbf{r}-\mathbf{r}')|^2\sum_\beta|\alpha_\beta|^2|\psi^{Cr}_{\beta}|^2d\mathbf{r'}\label{eq:STM}
\end{align}
where $|\alpha_\beta|^2=\frac{-1}{\pi}Im[Tr[G_{\beta,\beta}]]$ for $\beta\in\{xy,yz,zx,x^2-y^2,3z^2-r^2\}$ are the relative amplitudes of orbitals in the partial trace of the YSR states. We plot the simulated Cr $dI/dV$ map in the insets of Fig.\ref{Fig1} by convolving the partial trace amplitudes with Solid Spherical Harmonics with a Gaussian s-like STM tip wave function (see Supplemental Material).

However, for tips with atomic resolution, we expect to find a markedly different spatial distribution for the particle and hole contributions to the YSR states. Indeed, we show the spatial character of the particle and hole-like YSR states within the Bi surface layer in Fig. \ref{fig4}, for an energy $\omega=\pm 0.47$ meV. The $C_{4v}$ symmetry of the system is clearly recovered in the spatial distribution of the YSR state. Interestingly, we find that while the unoccupied YSR states are larger at the Cr atom, the occupied YSR states present a larger LDOS one lattice unit apart, maintaining the $C_{4v}$ symmetry. Overall, the particle component of the YSR state is more compact and closer to the Cr atom, while the hole part dominates further from the Cr atom.

This last behavior can be further studied by extending the study to other energies. To do this, we study the LDOS as a function of $\omega$ for several Bi atoms about the Cr atom.  In Fig. \ref{fig5}, we plot the Bi surface $p_z$ and $p_x+p_y$ partial traces as a function of distance from the first nearest neighbor Bi to the Cr impurity. At the Cr site, we find that the QP peaks that are clearly visible in the bulk traces, disappear, and the spectral weight of the QP peak is transferred to the YSR states in agreement with usual YSR calculations \cite{Yu_1965, Shiba_1968, Rusinov_1969}. The QP peaks recover as soon as the LDOS is evaluated away the Cr atoms, although they reflect the presence of the impurity for distances as large as 10 \AA.
In agreement with the spatial distribution in Fig. \ref{fig4}, we find that the prevalance of occupied or unoccupied peaks in the YSR state present a non-trivial pattern and the behavior can be inverted away from the impurity depending on the actual point on the surface. This distribution is a consequence of both the scattering potential of the Cr atom and the band structure of the superconducting substrate.

The orbital composition of the in-plane wave function has strong implications on the type and strength of hybridization experienced by nearby impurities. We expect this feature to play be important for understanding the emergence criteria for topological states in 1D spin-chains.

~
\section{Conclusion}
\label{sec:conclusions}

In this work, we have developed a comprehensive theoretical framework to describe impurity-induced in-gap states in superconductors, focusing on the case of a single Cr adatom on the Bi-terminated surface of $\beta$-Bi$_2$Pd. Our approach combines a realistic tight-binding model derived from first-principles calculations with a Green’s function-based scattering formalism within the Bogoliubov–de Gennes framework. This allows us to compute the full local density of states in the presence of a magnetic impurity without relying on diagonalization, preserving the full complexity of the superconducting substrate.

Our model captures key experimental observations, including the emergence of a single prominent Yu-Shiba-Rusinov (YSR) resonance with strong particle-hole asymmetry and the spatially isotropic $dI/dV$ maps near the Cr atom. By analyzing the orbital and spin decomposition of the local density of states, we identify the near-degeneracy of four symmetry-distinct YSR states corresponding to the $C_{4v}$ point group. This explains the absence of multiple resolved peaks, despite the multi-orbital character of the Cr impurity. The spin-resolved LDOS further reveals partial spin polarization due to the strong spin-orbit coupling of the Bi atoms, yielding a reduced spin contrast compared to simpler superconductors.

The spatial profiles of the YSR wavefunctions reveal distinct distributions for the particle and hole components, which may influence the effective exchange interaction between adatoms. These findings have important implications for the design and understanding of magnetic impurity chains on superconductors, particularly in the context of engineered topological states. Our formalism provides a robust and scalable platform for analyzing the role of atomic structure, orbital composition, and magnetic interactions in proximity-induced superconductivity.

Future extensions of this work may include self-consistent treatments of the order parameter suppression, multi-impurity systems with quantum spin dynamics, and the inclusion of unconventional pairing symmetries relevant to topological superconductors.

\section{Acknowledgments}
This material is  based on work primarily supported by the U.S. Department of Energy, Oﬃce of Science, Oﬃce of Basic Energy Sciences, under Award No. DE-SC0016379 (theory). For experimental results we thank projects  PID2021-127917NB-I00 by the Spanish Science Ministry program MCIN/AEI/10.13039/501100011033, QUAN-000021-01 by Gipuzkoa Provincial Council, IT-1527-22 and PIBA\_2024\_1\_0008 by the Basque Government, and the support of  the Institute for Basic Science (IBS-R027-D1). 
\bibliography{citations}

\end{document}